\documentclass[aps, pre, twocolumn, superscriptaddress, floatfix, notitlepage]{revtex4-2}

\usepackage{amsfonts, amsmath, amssymb, bbold, bigints, color, comment, enumitem, float, graphicx, epstopdf, lineno, makeidx, multirow, nicefrac, rotating, setspace, subfigure, url}
\usepackage[section]{placeins}
\usepackage[bookmarks, breaklinks, colorlinks]{hyperref}

\hypersetup{citecolor = blue, filecolor = black, linkcolor = red, urlcolor = blue}
\hypersetup{pdfauthor = {Dr. Chiranjit Mitra}}

\setlist[itemize,1]{label=$\bullet$}
\setlist[itemize,2]{label=$\circ$}
\setlist[itemize,3]{label=$\diamond$}
\setlist[itemize,4]{label=$\triangleright$}

\begin{document}

\title{Synchronization Dynamics of Heterogeneous, Collaborative Multi-Agent AI Systems}
\author{Chiranjit Mitra}
\email[Corresponding author: ]{chiranjitmitra4u@gmail.com}
\affiliation{Independent Researcher}

\begin{abstract}
We present a novel interdisciplinary framework that bridges synchronization theory and multi-agent AI systems by adapting the Kuramoto model to describe the collective dynamics of heterogeneous AI agents engaged in complex task execution.
By representing AI agents as coupled oscillators with both phase and amplitude dynamics, our model captures essential aspects of agent specialization, influence, and communication within networked systems.
We introduce an order parameter to quantify the degree of coordination and synchronization, providing insights into how coupling strength, agent diversity, and network topology impact emergent collective behavior.
Furthermore, we formalize a detailed correspondence between Chain-of-Thought prompting in AI reasoning and synchronization phenomena, unifying human-like iterative problem solving with emergent group intelligence.
Through extensive simulations on all-to-all and deterministic scale-free networks, we demonstrate that increased coupling promotes robust synchronization despite heterogeneous agent capabilities, reflecting realistic collaborative AI scenarios.
Our physics-informed approach establishes a rigorous mathematical foundation for designing, analyzing, and optimizing scalable, adaptive, and interpretable multi-agent AI systems.
This work opens pathways for principled orchestration of agentic AI and lays the groundwork for future incorporation of learning dynamics and adaptive network architectures to further enhance system resilience and efficiency.
\end{abstract}


\maketitle


\section{\label{sec:Introduction}Introduction}

The study of collective behaviour in complex networks has long been a cornerstone of applied mathematics, physics, biology and several other disciplines~\cite{strogatz2001exploring, albert2002statistical, dorogovtsev2002evolution, newman2003structure, boccaletti2006complex, newman2010networks}.
In computer science and artificial intelligence (AI), however, AI agents have only recently gathered attention whereby, enterprises are seeking to automate tasks via multiple agents interacting on a network, to accomplish a common goal~\cite{casper2025ai, bousetouane2025agentic, mukherjee2025agentic, okpala2025agentic}.
Historically, theories from physics have provided deep insights into AI and machine learning~\cite{hopfield1982neural, amit1987statistical, tkacik2009spin, raissi2019physics}.

In this paper, we seek to use this inspiration to propose a novel correspondence between collective behaviour (particularly, synchronization) and agentic AI applications~\cite{pikovsky2003synchronization}.
We adapt the Kuramoto model (KM), a paradigmatic framework in synchronization theory, to describe and analyze the collective behaviour of AI agents in collaborative tasks~\cite{rodrigues2016kuramoto}.
This innovative approach allows us to:
\begin{itemize}
\item represent the synchronization and coordination of AI agents working toward a common goal.
\item interpret the parallels between the model's components and dynamics with those of AI agents.
\item quantify and optimize agent interactions using established physics-based parameters.
\end{itemize}
By bridging the physics of complex systems and AI, we expect to open new avenues in designing efficient agentic AI systems.
This interdisciplinary approach leverages the rich theory and practice of collective behaviour and network theory, potentially aiding our understanding and implementation of multi-agent AI systems.


\subsection{\label{sec:Introduction_Motivation}Motivation}

\begin{figure}[H]
\begin{center}
\includegraphics[height=8.0cm, width=8.0cm]{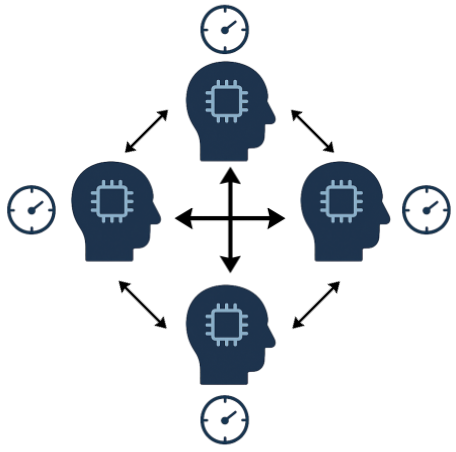}
\caption{\label{fig:Figure_1}(Color online) Illustration of a complete network of AI agents coordinating on a complex task, where each agent’s respective phase reflects its current position along its chain of thought.}
\end{center}
\end{figure}

Chain-of-Thought prompting has recently emerged as a groundbreaking concept in agentic AI, revolutionizing the way AI systems approach complex reasoning tasks~\cite{wei2022chain}.
This innovative technique guides AI models to break down intricate problems into sequential, logical steps, mirroring human-like thought processes.
By encouraging AI to articulate intermediate reasoning, Chain-of-Thought prompting significantly enhances problem-solving capabilities, improves decision-making processes, and increases the transparency of AI-generated outputs~\cite{kojima2022large}.
As visual articulation in this regard, Fig.~\ref{fig:Figure_1} provides a simple illustration of a complete network of AI agents coordinating on a complex task.

The dynamics of Chain-of-Thought (CoT) prompting techniques in AI and the evolution of agents in a Kuramoto-like system share intriguing parallels that can provide insights into complex problem-solving and collaborative dynamics.
The following is a correspondence between them:
\begin{itemize}
\item \textbf{Iterative Reasoning Process:}
    \begin{itemize}
    \item CoT: In Chain-of-Thought prompting, the model generates a series of intermediate steps, each building upon the previous one to reach a conclusion.
    \item KM: Agents in the system evolve their phases and amplitudes over time, with each iteration influenced by their previous state and the states of other agents.
    \end{itemize}
\item \textbf{Influence of Context:}
    \begin{itemize}
    \item CoT: Each step in the reasoning chain is influenced by the context provided by previous steps and the initial prompt.
    \item KM: The evolution of an agent's phase and amplitude is influenced by its own natural frequency and the states of neighbouring agents.
    \end{itemize}
\item \textbf{Emergent Coherence:}
    \begin{itemize}
    \item CoT: As the chain of thought progresses, a coherent line of reasoning emerges, leading to a more robust and justifiable conclusion.
    \item KM: As agents interact over time, their phases may synchronize, leading to a coherent collective behaviour represented by a high order parameter.
    \end{itemize}
\item \textbf{Adaptive Complexity:}
    \begin{itemize}
    \item CoT: The complexity of the reasoning chain can adapt to the difficulty of the problem, with more steps for more complex tasks.
    \item KM: The amplitude of agents can evolve to represent their relative importance or activity level in solving a complex task.
    \end{itemize}
\item \textbf{Convergence to Solution:}
    \begin{itemize}
    \item CoT: The reasoning process converges towards a final answer or solution to the posed problem.
    \item KM: The system may converge to a synchronized state, representing a collective solution or consensus.
    \end{itemize}
\item \textbf{Robustness to Perturbations:}
    \begin{itemize}
    \item CoT: A well-structured chain of thought can be more robust to small errors or uncertainties in individual steps.
    \item KM: A strongly coupled system of agents can maintain synchronization despite small perturbations to individual agents.
    \end{itemize}
\end{itemize}
By drawing these correspondences, we can see how the collaborative dynamics of agents in a Kuramoto-like system mirror the cognitive processes simulated by Chain-of-Thought prompting. This analogy suggests that principles from synchronization theory could potentially inform the development of more sophisticated prompting techniques, leading to more robust and coherent reasoning in AI systems.

\subsection{\label{sec:Introduction_Background}Background}

The past couple of years have witnessed major advancements at the confluence of AI agents research, mathematical modeling, and complexity science.
Recent literature now demonstrates a clear evolution from monolithic language models toward sophisticated multi-agent architectures - termed “agentic AI”, which harness coordinated networks of specialized agents to address complex challenges in reasoning, adaptation, and problem-solving~\cite{tran2025multi}.
The integration of rigorous mathematical frameworks and methodologies from network science is transforming our approach to the organization, analysis, and optimization of agent collectives~\cite{hu2024mathematical}.
Contemporary agentic systems employ advanced coordination strategies, such as distributed task allocation, hierarchical and peer-to-peer control, and consensus protocols, all grounded in principles from complexity theory and dynamical networks~\cite{xiao2025towards, agranat2025fueling}.
Recent surveys and technical analyses further underscore the importance of large language models (LLMs) as reasoning engines, facilitating enhanced autonomy, tool utilization, and dynamic memory within agentic workflows~\cite{wu2024autogen, ni2024mechagents}.

Importantly, emerging studies indicate that well-orchestrated agent collectives can surpass the capabilities of single-agent models, exhibiting phenomena such as collective intelligence, shared memory, and robust specialization through networked interaction~\cite{ni2024mechagents, tran2025multi}.
This convergence of AI and complexity science is exemplified by new frameworks that bridge macro-scale system architectures with local agent learning, leveraging both mathematical order parameters and scalable multi-agent coordination paradigms~\cite{hu2024mathematical, xiao2025towards}.
As research continues to incorporate advanced network topologies, adaptive interaction protocols, and distributed reasoning, the field is progressing towards a more principled and quantitative foundation for multi-agent AI, positioning it as a transformative paradigm for scientific inquiry and practical deployment~\cite{de2025open, sen2025modeling}.

Recent research further highlights the growing importance of context-aware and LLM-based multi-agent AI systems, which leverage coordinated networks of heterogeneous agents to autonomously tackle complex tasks with dynamic specialization and communication~\cite{du2024survey, xi2025rise, guo2024large, han2024llm}.
These advances closely align with the synchronization-based frameworks presented in this work, capturing key features such as agent influence, network effects, and scalable collaboration.
Additionally, the need for transparency and interpretability in agent interactions~\cite{chan2024visibility}, as well as the integration of multimodal information~\cite{durante2024agent}, emphasizes the relevance of adaptable, physics-informed models capable of representing and analyzing complex AI agent dynamics.
This convergence of modern AI research motivates the development of unified mathematical frameworks (such as ours) that support efficient coordination, adaptability, and interpretability in next-generation agentic AI systems.


\section{\label{sec:Methods}Methods}


\subsection{\label{sec:Methods_GD}General Dynamics}

In the following, we outline the general equations of motion for all nodes/agents of a networked dynamical system.
Consider a network of $N$ agents where the intrinsic dynamics of the $i\textsuperscript{th}$ agent (represented by the $d$-dimensional state vector $\mathbf{x}_{i}(t) = \left( x_{i}^{1},\, x_{i}^{2},\, \ldots,\, x_{i}^{d} \right)^{\text{T}}$) at time $t$ is described by:
\begin{widetext}
\begin{equation} \label{eq:DE_GD_Individual}
\dot{\mathbf{x}}_{i} = \mathbf{F}_{i} \left( \mathbf{x}_{i} \right);\, \mathbf{x}_{i} \in \mathbb{R}^{d};\, \mathbf{F}_{i}:\, \mathbb{R}^{d}\, \rightarrow\, \mathbb{R}^{d},\, \mathbf{F}_{i} = \left( F_{i}^{1} \left( \mathbf{x} \right),\, F_{i}^{2} \left( \mathbf{x} \right),\, \ldots,\, F_{i}^{d} \left( \mathbf{x} \right) \right)^{\text{T}};\, i = 1,\, 2,\, \ldots,\, N.
\end{equation}
\end{widetext}
The dynamical equations of the networked system read
\begin{equation} \label{eq:DE_GD_Network}
\dot{\mathbf{x}}_{i} = \mathbf{F}_{i} \left( \mathbf{x}_{i} \right) + \frac{\epsilon}{N} \sum\limits_{j = 1}^{N} A_{ij} \mathbf{H}_{ij} \left( \mathbf{x}_{i},\, \mathbf{x}_{j} \right),
\end{equation}
where:
\begin{itemize}
\item $\epsilon$ is the overall coupling strength.
\item $\mathbf{A}$ is the (directed) adjacency matrix which captures the interactions between the nodes such that $A_{ij} \neq 0$ if node $j$ influences node $i$.
\item $\mathbf{H}_{ij}\, \left[ \left( \mathbb{R}^{d},\, \mathbb{R}^{d} \right)\, \rightarrow\, \mathbb{R}^{d} \right]$ is an arbitrary coupling function from node $j$ to node $i$ such that $\mathbf{H}_{ij}$ and $\mathbf{H}_{ji}$ may be different, in general.
\end{itemize}
For the illustrations in this paper (Section \ref{sec:Results}), we consider non-identical nodal dynamics, symmetric adjacency matrices and non-identical coupling functions.


\subsection{\label{sec:Methods_KMD}Kuramoto Model Dynamics}

Let us consider a network of $N$ AI agents, each represented by a phase variable $\theta_{i}(t)$ and amplitude $r_{i}(t)$, where $i = 1,\, 2,\, \ldots,\, N$:
\begin{equation} \label{eq:DE_KM}
\begin{split}
\dot{\theta}_{i} = \omega_{i} + \frac{\epsilon}{N} \sum\limits_{j = 1}^{N} A_{ij} r_{j} \sin \left( \theta_{j} - \theta_{i} \right),\\
\dot{r}_{i} = r_{i} \left( \lambda - r_{i}^{2} \right) + \frac{\epsilon}{N} \sum\limits_{j = 1}^{N} A_{ij} r_{j} \cos \left( \theta_{j} - \theta_{i} \right),\\
\end{split}
\end{equation}
where:
\begin{itemize}
\item $\theta_{i}$ is the phase or \emph{state of progress} of agent $i$ (representing the progress of the task).
\item $r_{i}$ is the amplitude or \emph{strength} of agent $i$ (representing workload or importance).
\item $\omega_i$ is the natural frequency or \emph{inherent processing speed} of agent $i$.
\item $\lambda$ is a parameter controlling amplitude dynamics.
\item $\epsilon$ is the coupling strength between agents (representing inter-agent communication).
\item $A_{ij}$ is the adjacency matrix (representing the network connections between agents).
\end{itemize}
The natural frequency parameter $\left( \omega_{i} \right)$ not only reflects the intrinsic processing speed of each agent, but can also be interpreted as encoding fundamental aspects of the agent’s persona or operational profile, such as its preferred pacing, the domain of expertise, or the characteristic behavioral rhythm~\cite{lanham2025ai}.

In this model:
\begin{itemize}
\item the sine term in the equation promotes synchronization between connected agents, while the natural frequency term allows for individual differences in processing speed.
\item the phase equation now includes the amplitude $r_{j}$ in the coupling term, allowing stronger oscillators to have a greater influence on the phase dynamics.
\item the amplitude equation includes a term $r_{i} \left( \lambda - r_{i}^{2} \right)$ that governs the intrinsic amplitude dynamics.
\item the coupling in the amplitude equation uses a cosine function, complementing the sine function in the phase equation.
\end{itemize}
By analyzing the dynamics of this adapted Kuramoto model, we can gain insights into how the multi-agent AI system coordinates and progresses towards task completion.
The model allows us to study phenomena such as:
\begin{itemize}
\item emergence of synchronized sub-groups (specialized teams).
\item critical coupling strength for effective collaboration.
\item impact of network topology on task efficiency.
\item resilience to perturbations or agent failures.
\end{itemize}
This mathematical framework provides a foundation for designing and optimizing multi-agent AI systems for collaborative task completion.


\subsection{\label{sec:Methods_OP}Order Parameter}

To model task completion, we can introduce an order parameter $R(t)$ given by:
\begin{equation}
R(t) = \left| \frac{1}{N} \sum\limits_{j = 1}^{N} r_{j}(t) e^{i \theta_{j}(t)} \right|,
\end{equation}
where:
\begin{itemize}
\item $R(t)$ is the magnitude of synchronization at time $t$.
\item $N$ is the total number of agents.
\item $\theta_{j}(t)$ is the phase of agent $j$ at time $t$.
\item $r_{j}(t)$ is the amplitude of agent $j$ at time $t$.
\item $i$ is the imaginary unit.
\item $|\cdot|$ denotes the absolute value or magnitude.
\end{itemize}
The order parameter serves as a crucial metric for interpreting coordination among AI agents, particularly in terms of their synchronization and collective behaviour:
\begin{itemize}
\item the value of $R(t)$ ranges from 0 to 1.
\item when $R(t) = 0$, it indicates complete incoherence among the agents, meaning they are not synchronized at all.
\item conversely, when $R(t) = 1$, it signifies perfect synchronization (task completion), where all agents are moving in unison.
\item values between 0 and 1 indicate varying degrees of coordination.
\item for instance, an $R(t)$ value of 0.9 suggests that a large majority of agents are well-coordinated, while a value around 0.5 indicates moderate coordination, with some agents out-of-sync.
\end{itemize}

\begin{figure}[H]
\begin{center}
\includegraphics[height=8.0cm, width=8.0cm]{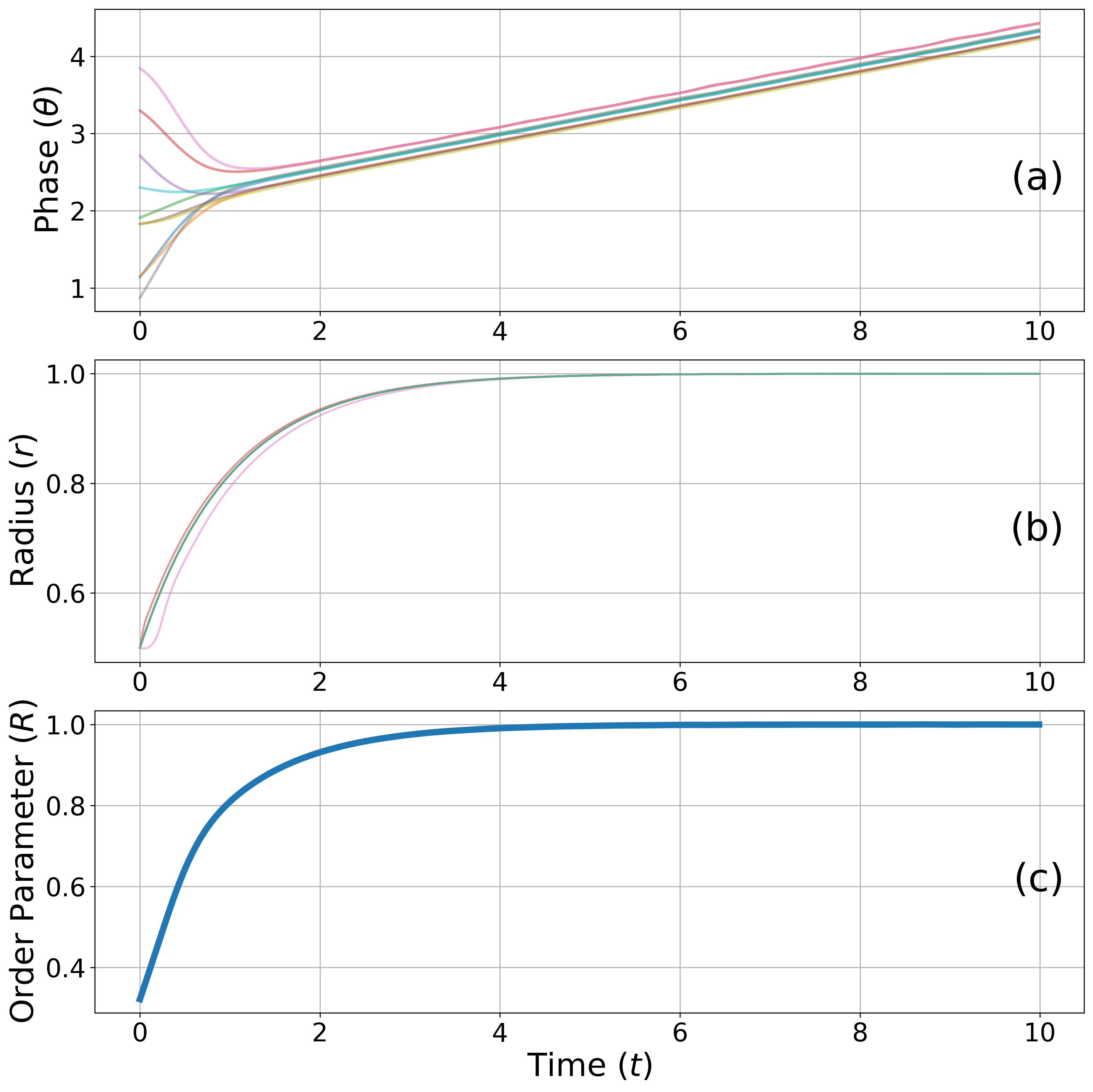}
\caption{\label{fig:Figure_2}(Color online) Temporal evolution of (a) phases. (b) radii. (c) order parameter for an all-to-all network of Kuramoto-like agents.}
\end{center}
\end{figure}


\subsection{\label{sec:Methods_HVHA}Homogeneous vs. Heterogeneous AI Agents}

Synchronization can indeed be achieved when all agents are identical, but this scenario is less interesting and realistic for multi-agent AI systems.
It does not capture the richness of real-world multi-agent systems where agents have different capabilities and specializations.
The more compelling case involves heterogeneous agents, where they synchronize despite diversity.


\subsection{\label{sec:Methods_TSA}Task-specific Adaptations}

Following task-specific adaptations are possible as further adjustments for the model in Eq.~(\ref{eq:DE_KM}):
\begin{itemize}
\item \emph{Sub-task allocation}: Divide the agents into sub-groups with different natural frequencies $\omega_{i}$ to represent specialization in sub-tasks.
\item \emph{Dynamic coupling}: Adjust $\epsilon$ based on task progress or difficulty, increasing coupling when more coordination is needed.
\item \emph{Network topology}: Design $A_{ij}$ to reflect the optimal communication structure for the task.
\item \emph{External influences}: Incorporate task-specific constraints, deadlines, or environmental factors.
\end{itemize}


\subsection{\label{sec:Methods_AE}Application Example: Modeling AI Agent Orchestration for HR Tasks}

To concretely illustrate how our model can be applied to multi-agent AI task orchestration, consider a company delegating HR-related tasks (such as processing job applications, handling payroll queries, and scheduling interviews) to a set of specialized AI agents.
Here, each agent represents a unit (e.g., recruiting, payroll, compliance) and possesses computational resources, such as \emph{number of available tokens} for LLM usage, \emph{share of compute} power allocated, and \emph{access to organizational data} for its operations~\cite{lanham2025ai}.

\begin{itemize}

\item \textbf{Resource Sharing:}
The amplitude $r_{i}$ can, in such a set-up, be mapped directly to the computational resources available to an agent $i$, for example, the number of tokens each agent is authorized to send to the central LLM service, or its share of cloud compute time.
When the total resource pool is constrained, $r_{i}$ naturally embodies both the capacity of an agent and its live resource budget.
The coupling structure $\left( \epsilon\, \text{and}\, A \right)$ can represent not only communication between agents but also explicit sharing or borrowing of tokens/compute among them, mirroring real-world resource dependencies.

\item \textbf{Dynamic Agent Orchestration:}
In agentic AI workflows, a central \emph{orchestrator} dynamically decides how many agents to instantiate, what their roles should be and how they should be interconnected to complete a task~\cite{sapkota2025ai, kim2025artificial}.
For example, if there is a surge in job applications, the orchestrator may increase the number of recruiting agents and adjust the network topology for more efficient internal communication.
Eq.~(\ref{eq:DE_KM}) and related dynamics allow us to \emph{quantitatively determine the minimal set and appropriate connections of agents} by optimizing coupling and resource assignment, ensuring that the emergent network achieves a high order parameter $R(t)$, i.e., effective task completion.

\item \textbf{Implications for System Design:}
Our model thus serves as a principled tool for:
    \begin{itemize}
    \item \emph{determining the number and specialization of AI agents} required for a given HR task, based on task load and current resource availability.
    \item \emph{configuring network topology} (e.g., all-to-all for rapid consensus, scale-free for hierarchical management) to best suit dynamic workloads.
    \item \emph{optimizing resource and token-sharing policies}, making sure that no key sub-task is starved of compute/tokens while maintaining global system efficiency.
    \item \emph{simulating system responses} to changes in workload, agent availability, or resource limitation, supporting robust and adaptive HR AI deployments.
    \end{itemize}

\end{itemize}
This approach generalizes to any business process where tasks, resource sharing, and adaptive agent configuration are critical for efficiency, providing both quantitative insight and actionable guidance for orchestrating agent networks in practice.


\subsection{\label{sec:Methods_AI_RD}Analytical Interpretation of the Radial Dynamics}

The radial dynamics in the model in Eq.~(\ref{eq:DE_KM}) with amplitude can be interpreted as follows:
\begin{itemize}
\item the term $r_{i} \left( \lambda - r_{i}^{2} \right)$ represents the intrinsic dynamics of the radius:
    \begin{itemize}
    \item $\lambda$ is a parameter that controls the intrinsic growth rate of the radius.
    \item the term $-r_{i}^{2}$ provides a non-linear saturation effect, which prevents unbounded growth.
    \end{itemize}
\end{itemize}
This equation has two main effects:
\begin{itemize}
\item when $r_i$ is small, the radius tends to grow $\left( \text{if}\, \lambda > 0 \right)$.
\item as $r_{i}$ increases, the negative $r_{i}^{2}$ term becomes dominant, causing the radius to decrease.
\end{itemize}
The equilibrium point for this intrinsic dynamics occurs when $r_{i} = \sqrt{\lambda}$.
In the context of AI agents:
\begin{itemize}
\item the radius $r_{i}$ can represent the \emph{influence} or \emph{activity level} of an agent.
\item $\lambda$ could represent the inherent capability or resources of the agent.
\item the saturation effect models limitations on an agent's growth or influence.
\end{itemize}
The coupling term $\frac{\epsilon}{N} \sum\limits_{j = 1}^{N} A_{ij} r_{j} \cos \left( \theta_{j} - \theta_{i} \right)$ can represent:
\begin{itemize}
\item resource sharing between agents.
\item mutual reinforcement of activity levels.
\item competition for limited resources.
\end{itemize}
This model allows for dynamic changes in agent importance or activity, which can be crucial in complex multi-agent AI systems where different agents may need to take on varying levels of responsibility depending on the task at hand.


\subsection{\label{sec:Methods_CI_RD}Contextual Interpretation of the Radial Dynamics for AI Agents}

The inclusion of a radius term in the Kuramoto model can be interpreted as representing the \emph{strength} or \emph{influence} of individual AI agents within a multi-agent AI system.
This interpretation allows for a more nuanced representation of agent interactions and their impact on the overall system dynamics.
In the following, we interpret the radius term in the context of AI agents working together:
\begin{itemize}
\item \textbf{Agent Influence}:
    \begin{itemize}
    \item The radius term can be seen as a measure of an agent's capability or effectiveness in contributing to the collective task.
    \item A larger radius would indicate a more influential or capable agent, while a smaller radius would represent a less impactful one.
    \end{itemize}
\item \textbf{Dynamic Adaptation}:
    \begin{itemize}
    \item \textbf{Learning and Improvement}: As agents improve their performance or acquire new skills, their radius (influence) may increase.
    \item \textbf{Resource Management}: The radius could reflect an agent's current resource allocation or energy level, fluctuating as resources are consumed or replenished.
    \item \textbf{Task Relevance}: An agent's radius might grow when its specialization is particularly relevant to the current sub-task and shrink when less relevant.
    \end{itemize}
\item \textbf{Interaction Dynamics}:
    \begin{itemize}
    \item \textbf{Weighted Contributions}: Agents with larger radii have a stronger effect on the phases of other agents, representing a form of weighted decision-making or influence on the collective behaviour.
    \item \textbf{Adaptive Coupling}: The overall coupling strength between agents becomes dynamic, potentially leading to more flexible and responsive collective behaviour, allowing the system to adapt to changing conditions or task requirements.
    \end{itemize}
\item \textbf{System-level Implications}:
    \begin{itemize}
    \item \textbf{Emergent Leadership}: Agents with consistently larger radii may naturally emerge as \emph{leaders} in the system, guiding the collective behaviour more strongly.
    This can result in a hierarchical structure within the multi-agent AI system, potentially improving coordination and decision-making.
    \item \textbf{Specialization}: The radius term could represent specialization in sub-tasks, with agents having larger radii in their areas of expertise.
    This allows the system to leverage the strengths of individual agents more effectively.
    \item \textbf{System Robustness}: The system may become more robust to individual agent failures, as the impact of low-radius (potentially malfunctioning) agents is naturally minimized.
    \end{itemize}
\end{itemize}
By incorporating the radius term, the model can capture more complex dynamics of AI agent collaboration, allowing for heterogeneous agent capabilities, adaptive influences, and emergent behaviours that more closely resemble real-world multi-agent AI systems.


\section{\label{sec:Results}Results}


\subsection{\label{sec:Results_A2L}All-to-all network of Kuramoto-like agents}

We simulate Eq.~(\ref{eq:DE_KM}) for $N = 10$ AI agents, with parameter values of $\lambda = 1.0$ and $\epsilon = 5.0$ \footnotemark[1].
We choose $\omega$ from a normal distribution with a mean $\mu = 0$ and standard deviation $\sigma = 0.5$, i.e., $\omega \sim \mathcal{N} \left( \mu = 0,\, \sigma = 0.5 \right)$.
As a specific network topology, we use an undirected all-to-all network with symmetric adjacency matrix $\left( A_{ij} = A_{ji} = 1,\, \forall\, i = 1,\, 2,\, \ldots,\, N \right)$.
We calculate and present the evolution of phases, radii and order parameter of the system in Fig.~\ref{fig:Figure_2}.

\begin{figure}[h]
\begin{center}
\includegraphics[height=5cm, width=8.0cm]{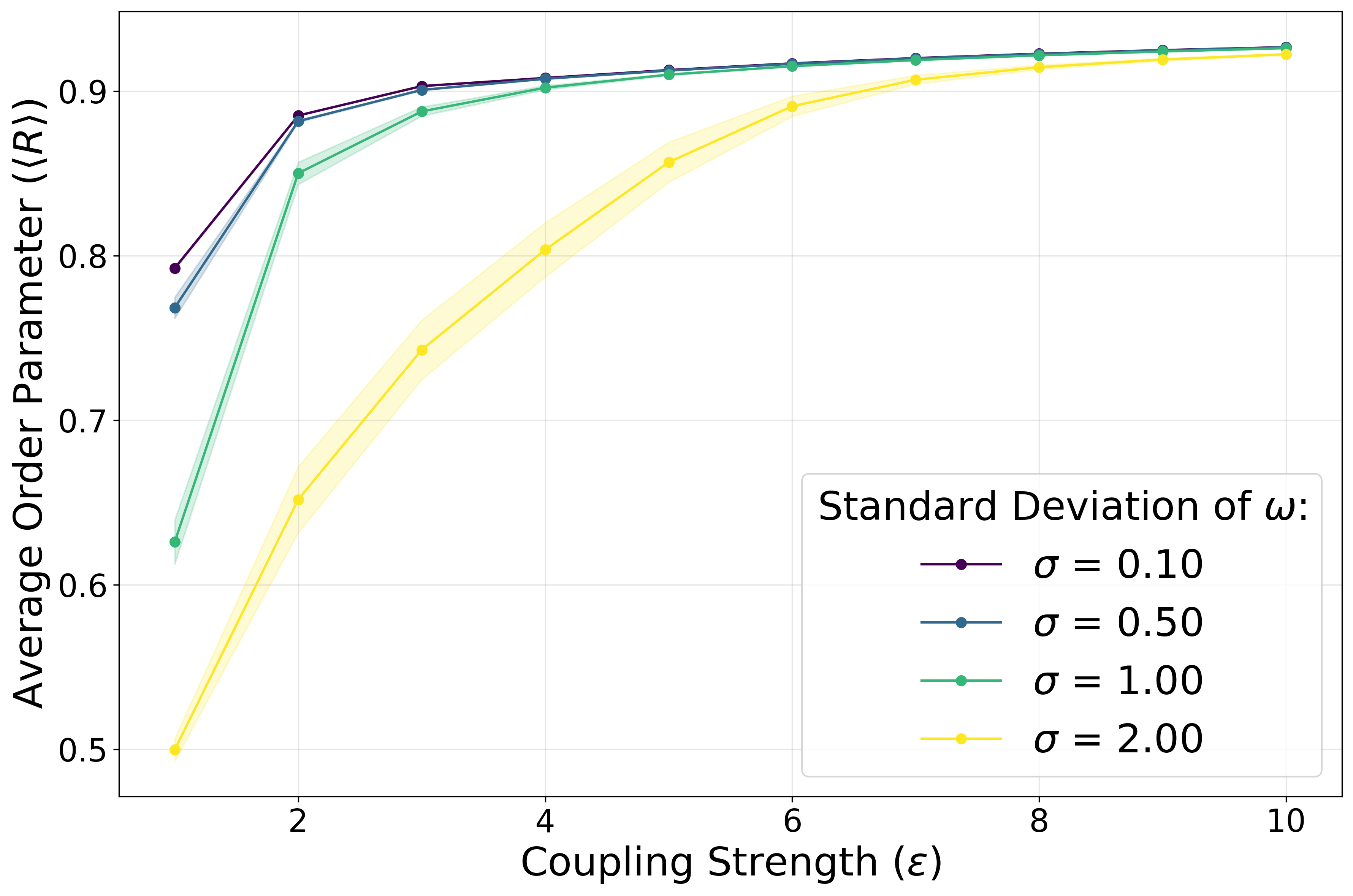}
\caption{\label{fig:Figure_3}(Color online) Relationship of the average order parameter $\left( \langle R \rangle \right)$ with the coupling strength $\left( \epsilon \right)$, for different values of standard deviation of the natural frequency of agents $\left( \sigma \right)$ interacting on an all-to-all network.}
\end{center}
\end{figure}

\footnotetext[1]{The code and data supporting this study are available on GitHub (subject to request and permission) at: \url{https://github.com/chiranjitmitra/sync_ai_agents}}

Interestingly, Fig.~\ref{fig:Figure_3} shows the relationship of the average order parameter $\left( \langle R \rangle \right)$ with the coupling strength $\left( \epsilon \right)$, for different values of standard deviation of the natural frequency of agents, $\sigma \in \{ 0.1,\, 0.5,\, 1.0,\, 2.0 \}$.
It clearly demonstrates that the network synchronizes better with increasing coupling strength despite heterogeneity between individual agents in the system.

\begin{figure}[h]
\begin{center}
\includegraphics[height=4.5cm, width=9.0cm]{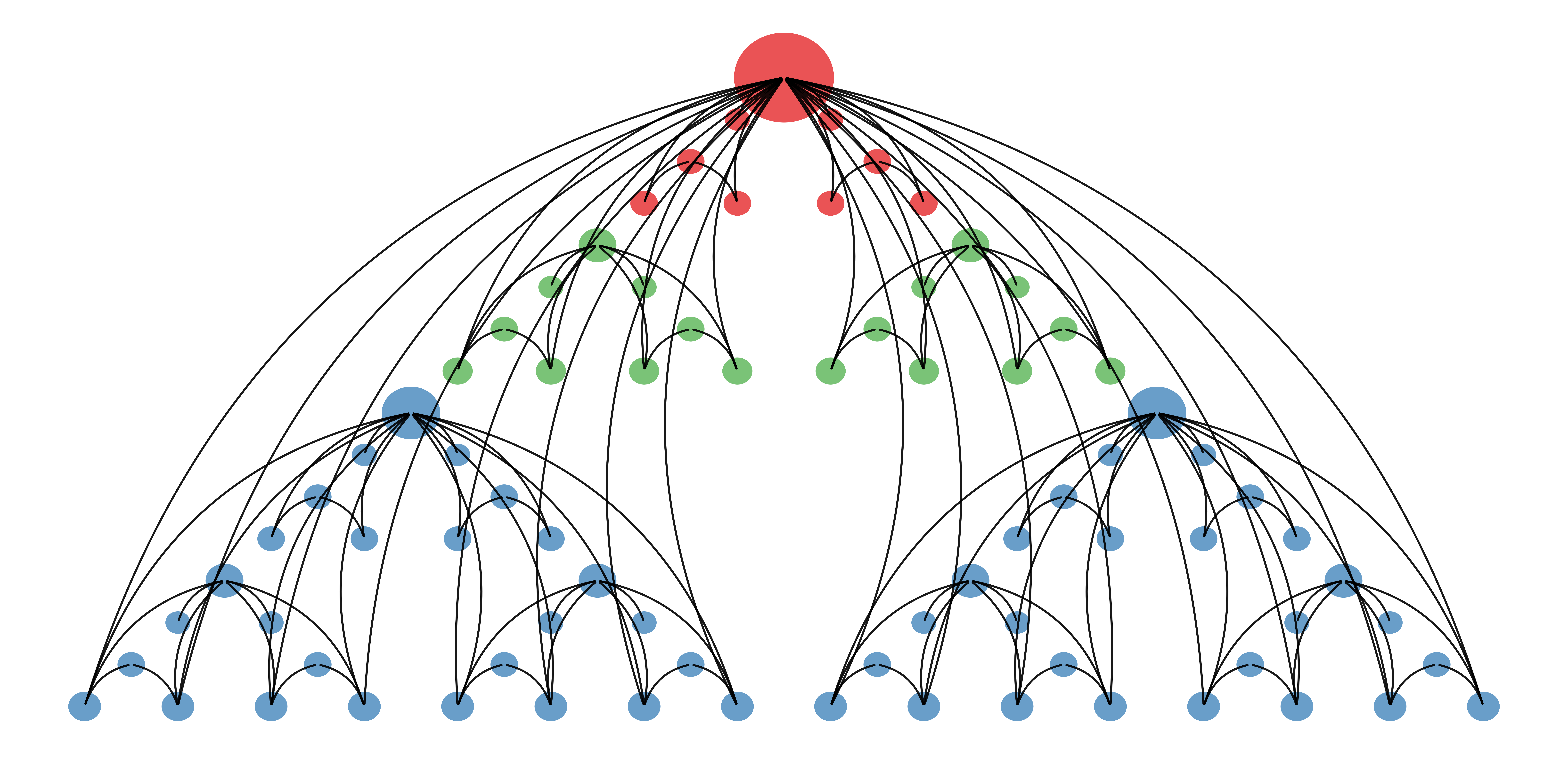}
\caption{\label{fig:Figure_4}(Color online) Network topology of the undirected deterministic scale-free network of $N = 81$ Kuramoto-like agents. The size of each node is proportional to its degree and the color (red, green and blue) indicates the level (first, second and third, respectively) of hierarchy to which the respective node belongs to.}
\end{center}
\end{figure}


\subsection{\label{sec:Results_DSF}Deterministic scale-free network of Kuramoto-like agents}

\begin{figure}[H]
\begin{center}
\includegraphics[height=8.0cm, width=8.0cm]{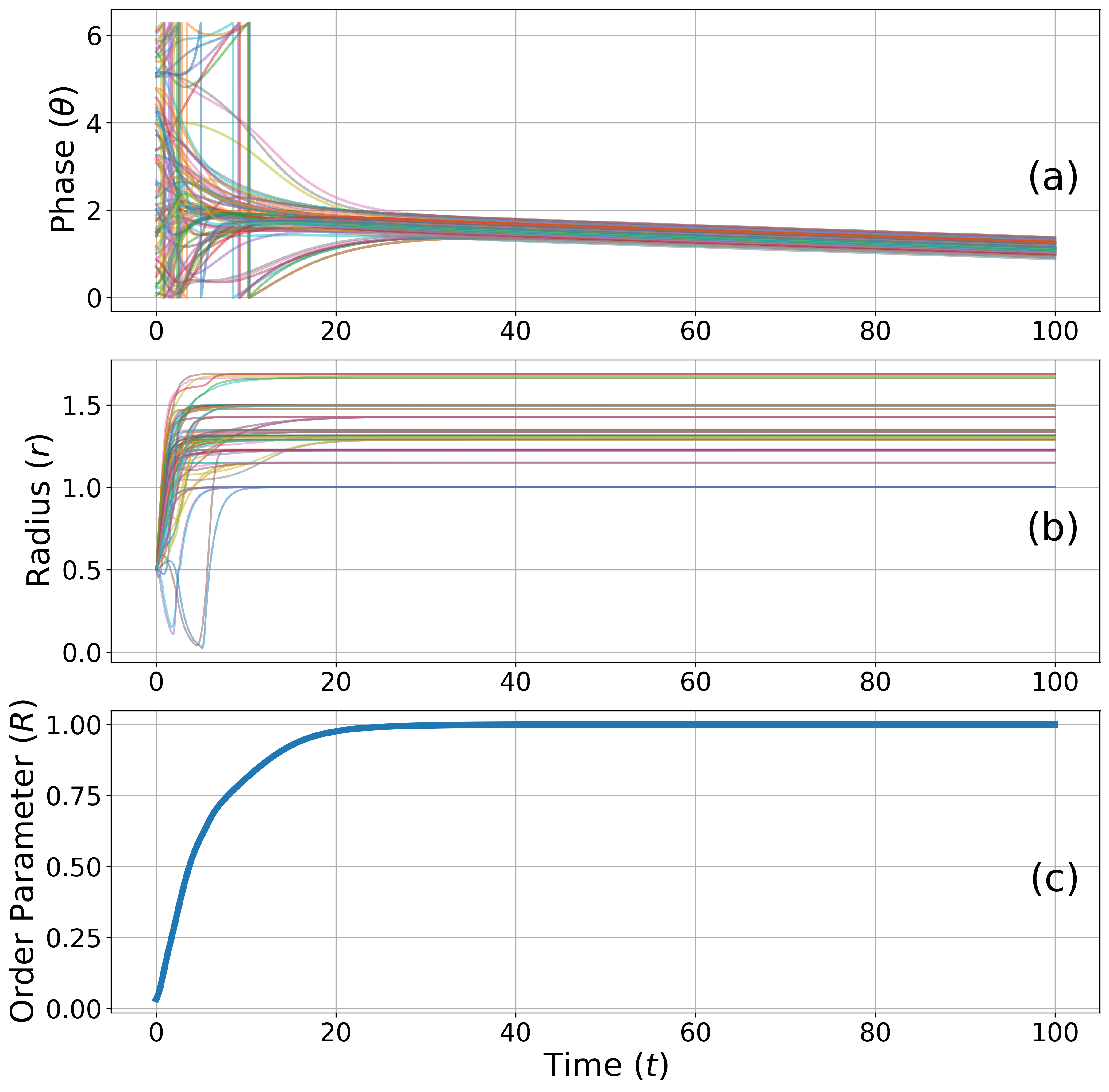}
\caption{\label{fig:Figure_5}(Color online) Temporal evolution of (a) phases. (b) radii. (c) order parameter for a determinstic scale-free network of Kuramoto-like agents.}
\end{center}
\end{figure}

For the simulations on the deterministic scale-free network, we consider a system of $N = 81$ agents governed by Eq.~(\ref{eq:DE_KM}), setting the parameters $\lambda = 1.0$ and $\epsilon = 30.0$.
The natural frequencies $\omega$ are sampled from a normal distribution with a mean $\mu = 0$ and a standard deviation $\sigma = 0.05$, i.e., $\omega \sim \mathcal{N} \left( \mu = 0,\, \sigma = 0.05 \right)$.
The communication structure corresponds to the undirected deterministic scale-free topology depicted in Fig.~\ref{fig:Figure_4}, characterized by a symmetric adjacency matrix ($A_{ij} = A_{ji} = 1$ if nodes $i$ and $j$ are connected and $A_{ij} = A_{ji} = 0$ otherwise)~\cite{barabasi2001deterministic, mitra2017multiple}.
Note that the choice of this topology is motivated by its correspondence with the hierarchical corporate structures which agentic AI networks often try to emulate.
Temporal dynamics of agent phases, amplitudes, and the associated order parameter are illustrated in Fig.~\ref{fig:Figure_5}.

\begin{figure}[H]
\begin{center}
\includegraphics[height=5cm, width=8.0cm]{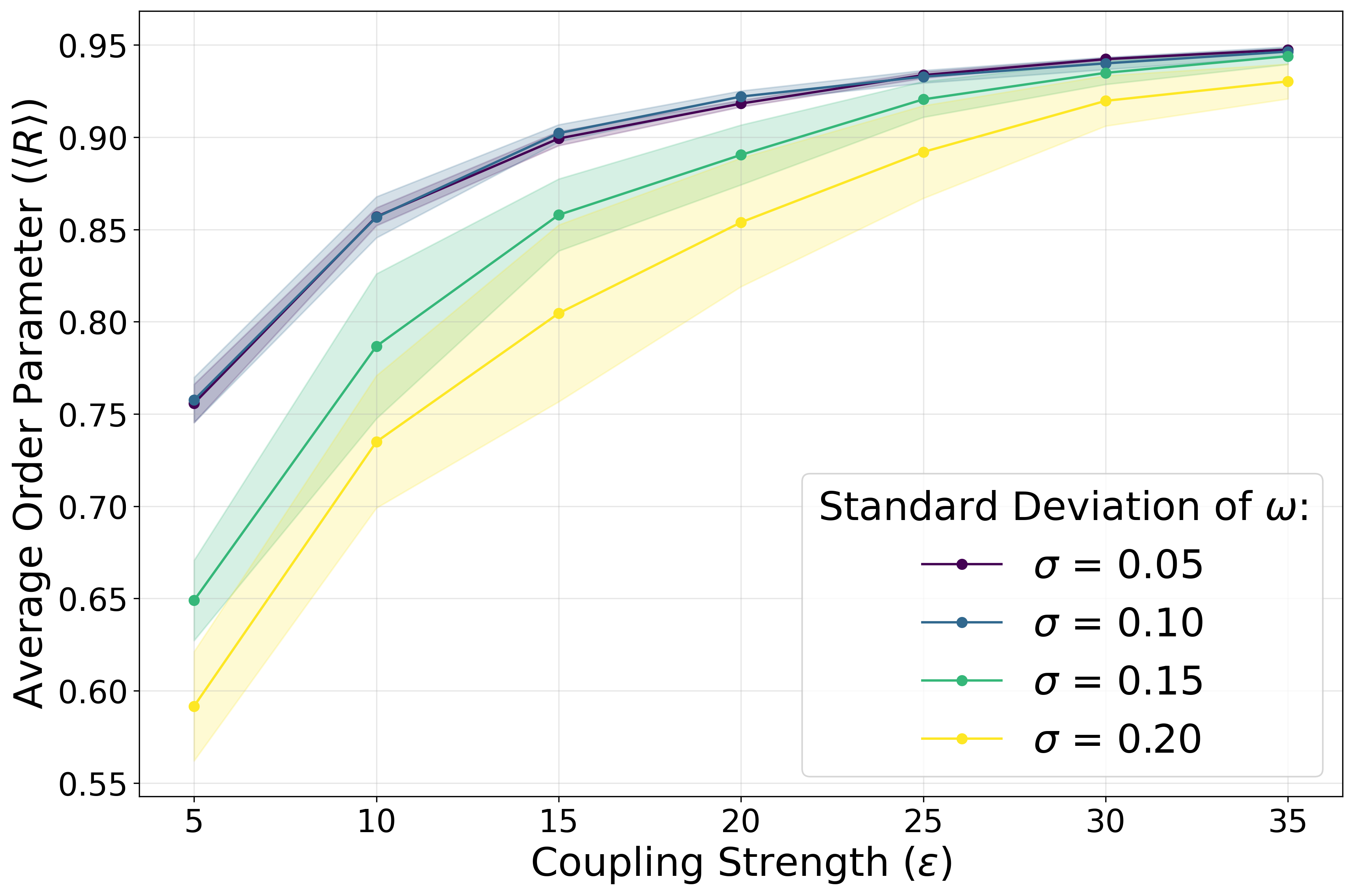}
\caption{\label{fig:Figure_6}(Color online) Relationship of the average order parameter $\left( \langle R \rangle \right)$ with the coupling strength $\left( \epsilon \right)$, for different values of standard deviation of the natural frequency of agents $\left( \sigma \right)$ interacting on a deterministic scale-free network.}
\end{center}
\end{figure}

Notably, Fig.~\ref{fig:Figure_6} illustrates how the average order parameter $\left( \langle R \rangle \right)$ depends on the coupling strength $\left( \epsilon \right)$ for various natural frequency dispersions, $\sigma \in \{ 0.05,\, 0.10,\, 0.15,\, 0.20 \}$.
These results highlight that even with pronounced heterogeneity among agents, the hierarchical scale-free network achieves enhanced synchronization as the coupling intensity increases.

\section{\label{sec:Conclusion}Conclusion}

In this paper, we have explored a novel synthesis between synchronization theory and multi-agent AI systems by adapting the Kuramoto model, a cornerstone in the study of collective dynamics, to the context of collaborative AI.
By framing AI agents as oscillators whose synchronization dynamics is governed by phase- and amplitude-interactions, we provide a rich theoretical foundation for analyzing and optimizing complex task execution by heterogeneous agent collectives.

Our approach demonstrates that the Kuramoto model can effectively represent the synchronization and coordination of AI agents working toward a common goal.
By drawing parallels between the iterative reasoning processes of Chain-of-Thought prompting and the model's dynamics, we have captured the nuances of agent influence and specialization, which are critical in real-world multi-agent AI systems.
The inclusion of amplitude dynamics allows differentiation of agent influence, dynamic adaptation to tasks, and robustness in the presence of agent diversity or failure.
The introduction of order parameters allows us to quantify group coordination, resilience to perturbations, and convergence to solutions, offering valuable insights into the efficiency and robustness of agent interactions.

Through simulations on both all-to-all and scale-free network topologies, it has been shown that increasing coupling strength enhances synchronization, even among highly heterogeneous agents.
This underscores the importance of effective communication and interaction strategies in multi-agent AI systems and reflects the model’s ability to capture real-world complexity.

This interdisciplinary approach potentially opens up multiple avenues of research.
The Kuramoto-inspired framework offers a unified, physics-informed basis for designing, monitoring, and controlling large-scale AI collaborations.
By mathematically identifying key drivers of collective behavior, such as coupling strength, agent diversity, and network topology, it facilitates systematic optimization and orchestration of multi-agent AI systems.
Furthermore, drawing on decades of insights from synchronization theory, this work invites a deeper interdisciplinary exploration into the principles of self-organization and emergent coherence, with the potential to inform the system design and development of more sophisticated and resilient next-generation AI systems capable of tackling complex, collaborative tasks.

In summary, this interdisciplinary approach of combining physics-based models with AI agent dynamics both enriches our understanding of collective behavior and equips practitioners with practical tools for engineering robust, scalable, and adaptive multi-agent AI systems.
Future research could further enhance these capabilities by incorporating learning dynamics, adaptive network structures, and more sophisticated agent models, paving the way for more resilient and capable agentic AI systems.


\begin{acknowledgments}
CM thanks his well-wishers for their support and encouragement.
Also, CM thanks \href{https://www.overleaf.com}{Overleaf} and \href{https://www.perplexity.ai}{Perplexity}.
\end{acknowledgments}


\bibliography{References}

\end{document}